\documentclass[sigconf, screen, nonacm]{acmart}

\setcopyright{none}
\AtBeginDocument{%
}
\acmISBN{978-1-4503-XXXX-X/2018/06}

\usepackage{amsmath,amsfonts}
\usepackage{algorithmic}
\usepackage{graphicx}
\usepackage{textcomp}
\usepackage{xcolor}
\usepackage{algorithm}
\usepackage{booktabs}
\usepackage{multirow}
\usepackage[utf8]{inputenc}
\usepackage{textgreek}
\usepackage{array}
\usepackage{tabularx}
\usepackage{colortbl}   % add after xcolor is loaded
\newcolumntype{Y}{>{\raggedright\arraybackslash}X}
\usepackage{threeparttable}
\usepackage[caption=false,font=footnotesize]{subfig}
\usepackage{pifont}

\usepackage{multicol}
\usepackage[normalem]{ulem}
\hypersetup{hidelinks}
\newcommand{\cOne}{\ding{172}}
\newcommand{\cTwo}{\ding{173}}
\newcommand{\cThree}{\ding{174}}
\newcommand{\mult}[1]{\,{\tiny\textit{$\times$#1}}}
\begin{document}

\title{Mutable Inference: Architecting LLM Accelerators for Inference-Time Learning}

\author{Yimin Gao}
\email{yg9bq@virginia.edu}
\affiliation{%
  \institution{University of Virginia}
  \city{Charlottesville}
  \state{Virginia}
  \country{USA}
}

\author{Liangtao Dai}
\email{kvf4sf@virginia.edu}
\affiliation{%
  \institution{University of Virginia}
  \city{Charlottesville}
  \state{Virginia}
  \country{USA}
}

\author{Melika Morsali}
\email{qfc2zn@virginia.edu}
\affiliation{%
  \institution{University of Virginia}
  \city{Charlottesville}
  \state{Virginia}
  \country{USA}
}

\author{Junting Huo}
\email{jaj8rm@virginia.edu}
\affiliation{%
  \institution{University of Virginia}
  \city{Charlottesville}
  \state{Virginia}
  \country{USA}
}

\author{Mircea Stan}
\email{mircea@virginia.edu}
\affiliation{%
  \institution{University of Virginia}
  \city{Charlottesville}
  \state{Virginia}
  \country{USA}
}

\renewcommand{\shortauthors}{Gao et al.}

\begin{abstract}
Long-context LLMs must retain growing histories, but full attention makes each new token increasingly expensive in computation and memory.
Inference-time learning (ITL) offers another approach. The model keeps training on the context it reads during inference, retaining what it learns in its weights for subsequent predictions. Recent work shows that ITL can match the quality of full-context attention %quality 
while limiting the actual use of attention to only a recent window of fixed size.
This bounds attention costs but adds %recurring 
\emph{Learning} (backward computation and weight updates) alongside 
\emph{Prediction} (the forward pass) costs. 
In contrast to ITL, \emph{inference accelerators} assume fixed weights, while \emph{training accelerators} target complete training steps rather than continuous learning during inference.

This work characterizes ITL on inference accelerators and identifies two opportunities. Attention's non-matrix operations leave enough matrix capacity idle during Prediction to cover %Learning's 
the matrix work needed for Learning, while selected layers can tolerate delayed %visibility of 
weight updates with little quality loss.
We exploit both with \emph{Learning While Serving (LWS)}, an inference accelerator architecture that shares compute hardware between latency-critical Prediction and deadline-flexible Learning.
Hardware-managed per-layer weight-versioning allows Learning to publish updates while in-flight computations continue reading older versions.
Resumable reduction blocks let long Learning kernels span fragmented idle windows and yield promptly to Prediction, while dedicated softmax-gradient units relieve contention on the shared vector unit.
On modeled datacenter and edge NPUs, LWS hides Learning entirely within Prediction, sustaining Prediction-only throughput while preserving 99.3\% of fully fresh Learning's quality benefit for TTT-E2E 3B at 128K tokens.
Across nine models, LWS achieves $1.6\times$ the throughput of serial execution and $1.3\times$ that of the best prior co-execution architecture.
RTL design and synthesis show that LWS adds 1.5\% to the NPU core area, without %LWS adds less than 1\% to die area, without %adding 
needing another MAC array.
By efficiently turning idle inference capacity into learning capacity, LWS establishes inference-time learning as a first-class workload for future LLM accelerators.
\end{abstract}

\maketitle

\section{Introduction}
\label{sec:intro}
Transformer-based Large Language Models (LLMs) have become remarkably capable, and they are increasingly asked to reason over long contexts, including long conversations, entire documents, codebases, and the accumulated histories of agentic systems~\cite{agent,agentmemory}. Full attention requires each new token to attend to all previous tokens. KV caching avoids recomputing past keys and values, but per-token attention computation and KV-cache  
% \Mircea{storage needs significanly} %still 
storage grow significantly with context length, making full-context attention a major bottleneck in long-context LLM inference.

Existing approaches 
% \Mircea{to mitigating the cost of KV caching} 
to reducing long-context inference costs differ in how they retain past context. KV cache quantization~\cite{KIVI,Hooper,GEAR} keeps every token in fewer bits, but the cache still grows linearly with context. Eviction~\cite{Scissorhands,H2O,SnapKV,StreamingLM} keeps only selected tokens and bounds the cache by discarding the rest. Recurrent models~\cite{Mamba2,DeltaNet} compress context into a fixed-size state, which fixes the per-token cost. 

\begin{figure*}[!tbp]
  \centering
  \newcommand{\panelh}{1.85in}%  raise until the row fills \textwidth
  \newcommand{\panelgap}{0.6em}%
  \begin{tabular}{@{}c@{}}
    {\small\textbf{(a) Inference-time learning}}\\[2pt]
    \includegraphics[height=\panelh]{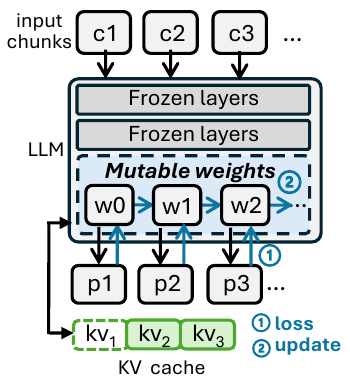}
  \end{tabular}\hspace{\panelgap}%
  \begin{tabular}{@{}c@{}}
    {\small\textbf{(b) Quality vs.\ input length}}\\[2pt]
    \includegraphics[height=\panelh]{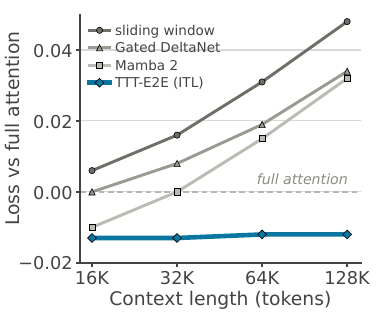}
  \end{tabular}\hspace{\panelgap}%
  \begin{tabular}{@{}c@{}}
    {\small\textbf{(c) Prefill cost vs.\ input length}}\\[2pt]
    \includegraphics[height=\panelh]{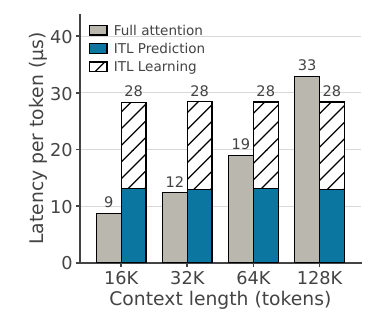}
  \end{tabular}
\caption{\textbf{Inference-time learning (ITL) trades growing attention costs for recurring Learning.}
(a) ITL learns from each input chunk, retaining information in mutable weights while keeping attention within a fixed window.
(b) TTT-E2E matches full-attention quality as context grows~\cite{ttte2e}.
(c) Unlike full attention, ITL keeps per-token prefill cost nearly constant, but adds recurring Learning that accounts for 54\% of its execution time (TTT-E2E 3B, B200).}
  \label{fig:teaser}
\end{figure*}

Inference-time learning (ITL)~\cite{ttte2e,TLM,LearnAtTestTime,Titans,TrainingDoneRight} instead retains context in a subset of the model weights. The model reads its context in chunks and continues training on each chunk during inference, so subsequent predictions benefit from what it has learned, while attention remains limited to only a recent window (Fig.~\ref{fig:teaser}(a)).
As shown in Fig.~\ref{fig:teaser}(b), TTT-E2E~\cite{ttte2e} maintains nearly constant loss relative to full attention as context grows to 128K tokens, while sliding-window attention and recurrent models such as Mamba-2 and Gated DeltaNet fall increasingly behind. ITL thus %trades 
replaces growing attention costs %for 
with the fixed cost of bounded attention and recurring Learning.

This recurring Learning changes how the model executes. During long-context prefill, ITL ingests the input in chunks, and each chunk involves two coupled phases. \emph{Prediction} performs the forward computation as in conventional inference, using the weights updated by Learning on preceding chunks, while \emph{Learning} computes gradients from the chunk and updates a subset of the weights.
For each mutable linear layer, Learning requires up to two additional matrix multiplications during backpropagation, one to propagate gradients to the preceding layer ($dX$) and another to compute gradients for updating the weights ($dW$).
As a result, the mutable weights are no longer fixed parameters loaded once and read repeatedly, but they become a \emph{state} that Learning continually updates and subsequent Prediction consumes.

This additional computation is %costly
not cheap. In the TTT-E2E reference implementation, each chunk's Prediction is followed by Learning, which usually must complete before the next chunk begins~\cite{ttte2e}.
At 128K tokens on an NVIDIA B200, ITL Prediction alone processes tokens $2.5\times$ faster than full attention, as shown in Fig.~\ref{fig:teaser}(c), since its attention is bounded to a fixed window.
However, Learning accounts for 54\% of ITL's total execution time, offsetting most of this speedup. Existing ML accelerators are not designed for %this 
such an execution pattern, in which backward computation and weight updates recur within inference. Inference accelerators~\cite{eyeriss} assume fixed weights, while training accelerators~\cite{tpu,Cambricon-Q} are optimized for large-batch training steps with weight updates only between steps. 

% A separate training datapath, however, may not be necessary. 
Yet much of Learning for ITL consists of matrix multiplications that inference accelerators already support. During Transformer inference, attention interleaves matrix operations with softmax, reductions, and other non-matrix operations, leaving the matrix array idle while dependent operations complete~\cite{Tandem,V10}. Our profiling on AWS Inferentia2 confirms this underutilization, showing that its tensor engine is active for only 27\% of a TTT-E2E forward layer. %To assess whether such idle capacity is sufficient for Learning, we model a TPUv5e-class datacenter NPU. On 
We assess the idle capacity for Prediction on a TPUv5e-class datacenter NPU and find out that, on this NPU, the matrix work for Learning for ITL would require only 33\% of the matrix work time left idle by Prediction. %61\% of the matrix work time left idle by Prediction. 
The challenge is therefore not insufficient aggregate matrix capacity, but how to exploit the available capacity without slowing down Prediction.

Our characterization identifies three insights that determine whether Learning can efficiently exploit this idle capacity.
\begin{enumerate}
    \item \textit{Weight-update visibility is temporally flexible.}
    Learning backpropagates in reverse layer order, so the weights Prediction needs first are updated last, which can stall Prediction if it must always use the latest weights. However, selected layers can consume weight versions one or two chunks old with little quality loss.

    \item \textit{Idle matrix capacity is fragmented.}
    The idle intervals during Prediction are often shorter than Learning's largest kernels.

    \item \textit{Learning creates a vector bottleneck.}
    Its attention and vocabulary softmax gradients drive vector demand to 2.1$\times$ the capacity that Prediction leaves idle on the modeled NPU.
\end{enumerate}

Building on these %findings
insights, we propose \emph{Learning While Serving (LWS)}, an inference accelerator architecture that supports latency-critical Prediction and deadline-flexible Learning on shared compute hardware. Rather than requiring Learning for each chunk %'s Learning 
to finish before the next Prediction proceeds, LWS exploits idle matrix capacity to execute Learning while enforcing per-layer weight-visibility deadlines. 

This paper makes the following contributions.

\begin{itemize}

\item \textbf{Architectural characterization of ITL.}
We characterize ITL on inference accelerators and identify three architectural insights for executing Learning alongside Prediction.

\item \textbf{LWS architecture.}
LWS introduces \cOne~hardware-managed per-layer weight versioning, which lets Learning publish new versions while Prediction continues reading the versions it acquired; \cTwo~resumable reduction blocks that let long Learning kernels span multiple idle windows and yield promptly to Prediction; and \cThree~dedicated softmax-gradient units that relieve the shared vector unit.

\item \textbf{Performance and hardware evaluation.}
For TTT-E2E 3B at 128K tokens, LWS matches Prediction-only throughput on modeled datacenter and edge NPUs while retaining 99.3\% of fully fresh Learning's NLL improvement, with only 1.5\% NPU core area overhead.
% , while adding less than 1\% to die area. 
Across nine models, LWS achieves $1.6\times$ the throughput of serial execution and $1.3\times$ that of the best prior co-execution architecture.
 
\end{itemize}

%\vspace{-5pt}
\section{Background and Motivation}
\label{sec:background}
\subsection{Execution of Inference-Time Learning}

\begin{figure}[!tbp]
  \centering
  \includegraphics[width=\linewidth]{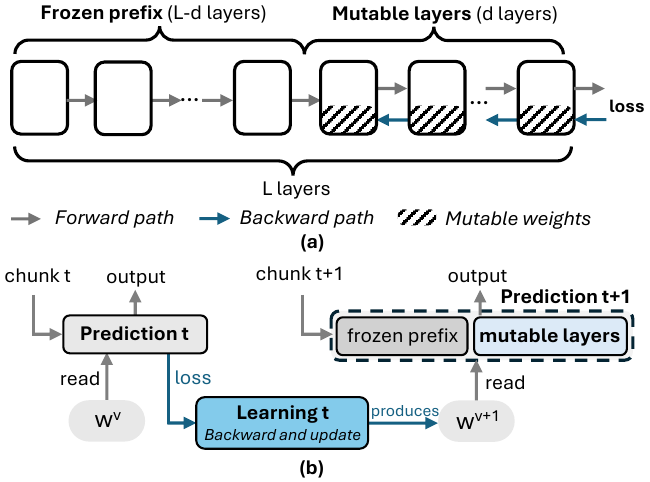}
  % \caption{\textbf{Execution dependencies in inference-time learning.} (a) Prediction traverses all $L$ layers, while Learning backpropagates through the $d$ mutable layers, updating selected weights. (b) Learning on chunk~$t$ produces updated weights that subsequent Prediction consumes upon reaching each mutable layer.}
\caption{\textbf{Execution dependencies in inference-time learning.}
(a) Prediction traverses all $L$ layers, while Learning updates the final $d$ mutable layers.
(b) Learning produces updated weights that subsequent Prediction consumes upon reaching the mutable layers.}
\label{fig:itl-exec}
\end{figure}

Fig.~\ref{fig:itl-exec}(a) illustrates how ITL processes a long input as a sequence of chunks, each containing a fixed number of consecutive tokens (e.g., 1,024 tokens in TTT-E2E 3B). For each chunk, Prediction passes through all $L$ Transformer layers to compute the next-token prediction loss. Learning then uses this loss to backpropagate through the final $d$ layers, updating selected mutable weights while the first $L-d$ layers remain frozen. This configuration is denoted as $\mathrm{ITL}(L,d)$, where $d$ is the number of layers traversed during backpropagation. For example, TTT-E2E 3B uses $\mathrm{ITL}(32,8)$, updating MLP weights in the final eight layers.

Fig.~\ref{fig:itl-exec}(b) shows the dependency between consecutive chunks. Prediction on chunk $t+1$ can begin processing the frozen prefix while Learning on chunk $t$ is still running. Under \emph{fully fresh execution}, Prediction must use the weights updated by Learning on chunk $t$ when it reaches each mutable layer. If that layer's weight updates are not yet complete, Prediction must stall until they become available. The timing of these weight dependencies creates opportunities to overlap Prediction and Learning, although incomplete updates can still delay Prediction.

For a linear layer $Y=XW$, where $X$ denotes the input activations, $W$ the weights, and $Y$ the output activations, backward propagation computes $dX=dY W^\top$ to propagate gradients to preceding layers and $dW=X^\top dY$ to compute gradients for updating the weights. Attention backward also involves matrix multiplications. Learning additionally requires elementwise operations and reductions, particularly for attention and vocabulary softmax gradients. Thus, Prediction and Learning both involve substantial matrix computation, although Learning introduces additional backward-specific operations.
\subsection{Untapped Capacity in Inference Accelerators}
\label{sec:char-util}

Modern inference accelerators combine matrix compute arrays with vector and special-function units (SFUs), on-chip buffers, and off-chip memory interfaces~\cite{TPUv4,TPUv5e,AWSNeuron}. In Transformer inference, projections and feed-forward networks involve substantial matrix computation that maps naturally onto the matrix datapath. Attention, however, interleaves matrix multiplications with frequent vector/scalar operations, such as scaling, exponentiation, and reductions for softmax~\cite{Tandem,SystolicAttention}. Since subsequent matrix operations must wait for these intermediate results, the matrix array can remain idle while other units are active~\cite{Tandem,V10,SystolicAttention}.

\begin{figure}[!tbp]
    \centering
    \includegraphics[width=\columnwidth]{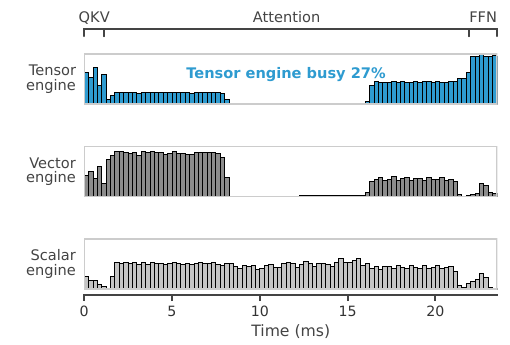}
\caption{\textbf{Matrix-compute headroom during inference.} Measured engine activity for one TTT-E2E forward layer on AWS Inferentia2 (Neuron compiler, default materialized attention). During attention, the tensor engine waits for scalar operations, remaining active for only 27\% of the overall layer.}
    \label{fig:resource-util}
\end{figure}

% We observe this underutilization on two commercial inference accelerators.
% Fig.~\ref{fig:resource-util} shows measured engine activity for a TTT-E2E forward layer on the AWS Inferentia2 NPU, which contains dedicated tensor, vector, and scalar engines~\cite{AWSNeuron}. During attention, scalar operations occupy substantial execution time, leaving the tensor engine idle while waiting for dependent results. Across the measured forward layer, the tensor engine is active for only 27\% of execution time. 
% Our profiling on TPU v5e~\cite{TPUv5e} additionally shows that the TTT-E2E forward pass achieves only 19\% of peak matrix throughput, further confirming the underutilization of matrix compute resources, consistent with substantial underutilization of matrix compute resources.
We observe this underutilization on two commercial inference accelerators. Fig.~\ref{fig:resource-util} shows measured tensor, vector, and scalar engine activity for a TTT-E2E forward layer on AWS Inferentia2~\cite{AWSNeuron}. During attention, the tensor engine frequently waits for scalar operations to complete, leaving substantial gaps between matrix computations. Across the measured forward layer, the tensor engine is active for only 27\% of execution time. Profiling on TPU v5e~\cite{TPUv5e} similarly shows that TTT-E2E Prediction achieves only 19\% of peak matrix throughput. These measurements indicate substantial unused matrix capacity during inference, providing an opportunity to execute additional computation on the existing hardware.

Together, these observations suggest an opportunity to use existing matrix resources for Learning without adding a separate MAC array. 
This motivates rethinking how inference-time learning can be mapped onto inference accelerators, particularly \textbf{how Learning can exploit underutilized compute resources without sacrificing Prediction throughput}. Section~\ref{sec:characterization} explores the execution dependencies and hardware resource constraints that shape this opportunity.

\section{Architectural Characterization of Inference-Time Learning}
\label{sec:characterization}

% This section explores three aspects of ITL execution on inference hardware: weight-update visibility, the timing of idle compute windows, and the resource demands of backward propagation. Together, they reveal opportunities and constraints for executing Learning alongside Prediction.

This section characterizes three key aspects of ITL execution: weight-update visibility, fragmented idle compute windows, and backward propagation's resource demands.

%\vspace{-5pt}
\subsection{Reverse-Order Updates Expose Temporal Flexibility}
\label{sec:char-visibility}

Learning and Prediction traverse mutable layers in opposite orders. During backpropagation, gradients for the output-most mutable layer become available first, while those for the input-most layer become available last. The next chunk's Prediction reaches these layers in the reverse order, requiring the input-most layer's weights first. Under \emph{fully fresh execution}, where Prediction always uses the latest updated weights, this ordering can stall Prediction even when some updates are already available.

However, updated weights do not necessarily need to become visible to Prediction immediately. 
We evaluate this flexibility by delaying when Prediction uses each mutable layer's updated weights, while Learning continues to update them every chunk.
A lag of $\ell$ chunks means that Prediction consumes an updated weight version $\ell$ chunks later than under fully fresh execution. 
Learning computes each chunk's gradients using the same weight versions as the corresponding Prediction, then applies the updates to the latest master weights.
Table~\ref{tab:lag-quality} shows that Learning retains most of its quality benefit with delayed weight visibility. For TTT-E2E at 128K tokens, fully fresh Learning reduces perplexity from 11.26 without weight updates to 10.87, with both cases using the same bounded sliding-window attention.
Retention measures the fraction of the corresponding NLL improvement preserved under a given policy. In these quality experiments, each layer uses a fixed lag equal to the maximum permitted by that policy.
Delaying weight visibility by one chunk for all mutable layers retains 96.3\% of the benefit.

The reverse traversal of Learning and Prediction motivates assigning different visibility lags across mutable layers. Input-side layers receive their gradients last but are needed first by Prediction, making their updates more likely to stall forward progress. Allowing longer visibility lags for these layers can relax this dependency, while output-side layers can remain fresher because their gradients become available earlier and Prediction reaches them later. 
Following this reasoning, the mixed-lag policy \texttt{1111|0000}, ordered from input-most to output-most mutable layer, retains 99.3\% of Learning's benefit at 128K, outperforming uniform lag 1 across all three evaluated context lengths. In contrast, freezing the four input-most mutable layers retains only 76.2\% of the benefit.

\textit{Insight 1.}
Weight-update visibility is temporally flexible and varies across layers. Delaying selected updates preserves nearly all of Learning's quality benefit, so each layer's update can follow its own visibility deadline rather than becoming visible to the next Prediction immediately.

% \begin{table}[!tbp]
%     \centering
% \caption{\textbf{Quality retention under weight-version lag.}
% Retained share of fully fresh ITL's NLL benefit (TTT-E2E 3B). Digits are per-layer lags (M1 first), and F is frozen. $\bar{\ell}$ denotes the mean lag across M1--M8.}
%     \label{tab:lag-quality}
%     % \footnotesize
%     \setlength{\tabcolsep}{3pt}
%     \begin{tabular}{@{}lccccc@{}}
%         \toprule
%         Policy & Lag M1--M8 & $\bar{\ell}$ & 32K & 64K & 128K \\
%         \midrule
%         Fully fresh       & \texttt{0000|0000} & 0     & 100   & 100   & 100   \\
%         Uniform lag 1     & \texttt{1111|1111} & 1.0   & 92.94 & 95.43 & 96.29 \\
%         \textbf{Mixed lag} & \texttt{2111|1000} & 0.75  & \textbf{95.49} & \textbf{97.59} & \textbf{98.57} \\
%         Freeze input half & \texttt{FFFF|0000} & --    & 69.07 & 72.75 & 76.23 \\
%         \bottomrule
%     \end{tabular}
% \end{table}

\begin{table}[!tbp]
    \centering
\caption{\textbf{Quality retention under weight-version lag.}
Retained share of fully fresh ITL's NLL benefit (TTT-E2E 3B). Digits are per-layer lags (M1 first), and F is frozen. $\bar{\ell}$ denotes the mean lag across M1--M8.}
    \label{tab:lag-quality}
    % \footnotesize
    \setlength{\tabcolsep}{3pt}
    \begin{tabular}{@{}lccccc@{}}
        \toprule
        Policy & Lag M1--M8 & $\bar{\ell}$ & 32K & 64K & 128K \\
        \midrule
        Fully fresh       & \texttt{0000|0000} & 0     & 100   & 100   & 100   \\
        Uniform lag 1     & \texttt{1111|1111} & 1.0   & 92.94 & 95.43 & 96.29 \\
        \textbf{Mixed lag} & \texttt{1111|0000} & 0.5  & \textbf{97.60} & \textbf{98.72} & \textbf{99.34} \\
        \textbf{Mixed lag} & \texttt{2111|1000} & 0.75  & \textbf{95.49} & \textbf{97.59} & \textbf{98.57} \\
        Freeze input half & \texttt{FFFF|0000} & --    & 69.07 & 72.75 & 76.23 \\
        \bottomrule
    \end{tabular}
\end{table}
% The distinction between delayed and omitted updates is substantial. At 128K tokens, freezing the four input-most mutable layers retains only 76.2\% of Learning's benefit. Delaying update visibility preserves the learning process, whereas freezing a layer removes its updates entirely.

% \textit{Insight 1.}
% Weight-update visibility is temporally flexible. \edit{Asynchronous and pipelined training have long shown that updates computed on slightly stale weights still converge~\cite{hogwild,pipedream2bw}. In ITL, this tolerance differs across layers, so the visibility of each layer can be delayed according to when Learning's reverse traversal reaches it.} Delaying the visibility of selected layers' updates preserves nearly all of Learning's quality benefit, whereas omitting updates sacrifices substantially more of it.
% \textit{Insight 1.}
% Weight-update visibility is temporally flexible. Delaying selected updates preserves nearly all of Learning's quality benefit, allowing each layer's update to follow its own visibility deadline rather than requiring immediate use by the next chunk.

\subsection{Learning Kernels Outlast Idle Windows}
\label{sec:char-fragmentation}

Although attention computation leaves substantial matrix capacity unused on inference accelerators, this idle time is not continuously available. Instead, attention interleaves matrix multiplications with vector and scalar operations, creating idle periods of varying lengths on the matrix array.
As shown in Fig.~\ref{fig:resource-util}, on AWS Inferentia2 with the Neuron compiler's default materialized attention, the tensor engine has one relatively long idle interval during softmax, but otherwise idles only briefly between operations.
As a result, executing Learning kernels to completion during an idle period could delay Prediction when it needs the matrix array again.

Modern inference implementations increasingly use FlashAttention-style tiled attention~\cite{dao2022flashattention} to reduce memory traffic. Conventional attention computes $O=\mathrm{softmax}(QK^\top/\sqrt{d_k})V$ in successive stages. Since softmax requires attention scores across all keys, a conventional implementation computes and stores the full score matrix before applying softmax and multiplying the resulting probabilities by $V$. These intermediate matrices can introduce substantial off-chip memory traffic. FlashAttention instead partitions the computation into smaller tiles that fit in on-chip memory, using incremental softmax to interleave matrix multiplications and normalization without storing the full intermediate matrices. 
This substantially reduces memory traffic and improves attention efficiency. However, the finer-grained interleaving of matrix and softmax operations further fragments the matrix array's idle time, creating shorter, more frequent windows in which Learning can execute without delaying Prediction. 
Despite this fragmentation, the substantial idle capacity remains. The matrix array on Inferentia2's NeuronCore-v2 is idle for more than half of FlashAttention's execution time~\cite{SystolicAttention}.

\textit{Insight 2.}
Aggregate idle matrix capacity does not guarantee effective overlap. Learning's long kernels may exceed individual idle windows, despite their total computation fitting within Prediction's unused capacity.

\subsection{Softmax Gradients Expose a Vector Bottleneck}

\begin{figure}[!tbp]
  \centering
  \includegraphics[width=.83\columnwidth]{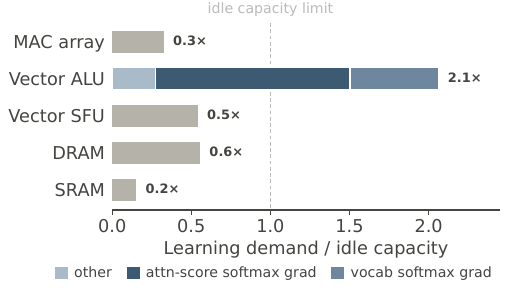}
  \caption{\textbf{Learning demand versus idle inference capacity.}
  Resource demand normalized to Prediction's unused execution
  or bandwidth capacity on a modeled TPUv5e-class NPU.
  % (394 TOPS INT8, 32 MB SRAM, 819 GB/s HBM).
  Most resources have sufficient aggregate capacity,
  while softmax gradients drive vector demand to
  2.1$\times$ its idle capacity.}
  \label{fig:characterization-compute}
\end{figure}

Reusing idle matrix capacity also requires sufficient support from other hardware resources. Although Learning's matrix operations can use the existing MAC arrays, backward propagation also involves non-matrix operations that compete with Prediction for the vector unit. We compare Learning's resource demands against Prediction's unused capacity on a modeled TPUv5e-class inference NPU. Fig.~\ref{fig:characterization-compute} shows that Learning's matrix and memory demands fit within the available aggregate capacity, but its vector demand reaches 2.1$\times$ the unused vector capacity.

\begin{figure*}[!tbp]
    \centering
    \includegraphics[width=\textwidth]{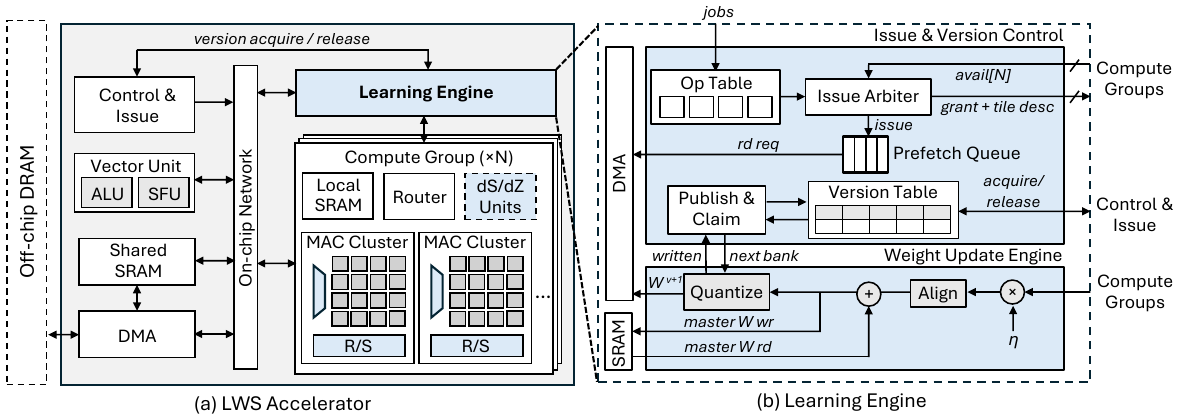}
    \caption{\textbf{LWS accelerator architecture.} (a) LWS extends an inference NPU with a Learning Engine and lightweight compute-group modifications while retaining the existing matrix arrays and memory hierarchy. Dashed units exist only in eight compute groups. (b) The Learning Engine coordinates task issue, weight versioning, and update publication.}
    \label{fig:lws-arch}
\end{figure*}

Most of this vector demand comes from two operations, the attention softmax gradient and the vocabulary softmax gradient. Both involve elementwise computations and reductions across attention positions or vocabulary entries, requiring substantial work on the shared vector unit. Together, they account for 86.8\% of Learning's vector work. Their results are also needed by subsequent backward computations, so contention on the vector unit can delay Learning's matrix operations despite available matrix capacity.

% To understand how much these two operations contribute to the bottleneck, we examine Learning's remaining vector demand when their contributions are excluded. Excluding the attention softmax gradient alone reduces demand to 0.8$\times$ the unused capacity, while excluding the vocabulary softmax gradient alone still leaves 1.5$\times$. Excluding both reduces the remaining demand from 2.1$\times$ to 0.3$\times$.
% FlashAttention-2~\cite{flashattention2} simplifies the attention softmax gradient by replacing a reduction across attention positions with a dot product between the attention output and its gradient. Even with this optimization, Learning's vector demand remains at 1.5$\times$ the unused capacity, as the vocabulary softmax gradient remains a major contributor.
FlashAttention~\cite{dao2022flashattention, flashattention2} simplifies the attention softmax gradient by replacing a reduction across attention positions with a dot product between the attention output and its gradient. Even with this optimization, the remaining elementwise operations still require substantial vector computation. 
To quantify their contribution to the bottleneck, we exclude the two softmax gradients from the resource accounting. Excluding both reduces Learning's vector demand from 2.1$\times$ to 0.3$\times$ the unused capacity, while excluding only the attention or vocabulary gradient reduces demand to 0.8$\times$ or 1.5$\times$, respectively.

% To understand how much the two softmax gradients contribute to the bottleneck, their vector demands are excluded individually. Excluding the attention softmax gradient reduces Learning's vector demand from 2.1$\times$ to 0.8$\times$ the unused capacity, while excluding the vocabulary softmax gradient reduces it to 1.5$\times$. Excluding both leaves only 0.3$\times$ the unused capacity. These results show that both operations contribute substantially to the vector bottleneck even with FlashAttention-2 optimization.

\textit{Insight 3.}
Attention and vocabulary softmax gradients account for most of Learning's vector demand, creating a resource bottleneck that limits overlap with Prediction despite sufficient aggregate matrix capacity.

% \begin{table}[!tbp]
% \centering
% \caption{\textbf{Mapping Transformer operations onto a conventional inference NPU.} Highlighted gradients are unique to Learning and execute on the vector unit.}
% \label{tab:learning-ops}
% \footnotesize
% \setlength{\tabcolsep}{4pt}
% \begin{tabular}{@{}lll@{}}
% \toprule
% Phase & Operation & Hardware \\
% \midrule
% Prediction & Linear projections & MAC array \\
%            & Attention matmuls & MAC array \\
%            & Attention softmax & Vector unit \\
%            & Vocabulary logits & MAC array \\
% \midrule
% Learning   & $dX$, $dW$ & MAC array \\
%            & $dQ$, $dK$, $dV$, $dP$ & MAC array \\
%            & \textbf{Attention softmax grad. ($dS$)}
%            & \textbf{Vector unit} \\
%            & \textbf{Vocabulary softmax grad. ($dZ$)}
%            & \textbf{Vector unit} \\
%            & Other backward operations & Vector/SFU \\
% \bottomrule
% \end{tabular}
% \end{table}

% \vspace{-10pt}
\section{LWS Architecture}
\label{sec:architecture}

\subsection{Architecture Overview}
\label{sec:arch-overview}
LWS extends an inference NPU to support both Prediction and Learning on shared MAC arrays, without introducing a separate matrix-compute engine. Prediction retains priority over the shared hardware, while Learning uses otherwise idle matrix capacity and yields promptly when Prediction needs the array. 
Resumable execution allows Learning kernels to span multiple idle periods, while hardware-managed per-layer weight versioning allows Prediction to proceed using previously acquired versions as Learning produces updates. Dedicated softmax-gradient units further reduce contention for the shared vector resources.
Together, these mechanisms allow Learning to progress alongside Prediction while meeting each layer's weight-visibility deadline. 

% Fig.~\ref{fig:lws-arch} illustrates the overall architecture. LWS builds on an inference accelerator with compute throughput and memory bandwidth matched to TPU v5e~\cite{TPUv5e}. 
% The matrix datapath consists of 145 compute groups, each with four MAC clusters, a router, and local SRAM. Each MAC cluster contains 16 MXINT8 MAC tiles. The compute groups also access a shared vector unit and chip-level SRAM.
% Unlike TPU v5e's four large Matrix Multiplication Units (MXUs), our NPU organizes comparable aggregate compute capacity into 580 independently schedulable parallel MAC clusters, enabling finer-grained sharing between Prediction and Learning. 
% LWS retains the existing compute-group structure and adds a \emph{Learning Engine} to manage Learning tasks, weight versions, and weight updates.
% Each compute group contains the same MAC clusters, with lightweight resumption and suspension (R/S) support added to every cluster so that Learning kernels can continue across multiple idle intervals. A subset of compute groups also includes specialized dS/dZ units for softmax gradients and a Stash Quantization (Stash Q) unit for preparing stashed activations for weight-gradient computation (Section~\ref{sec:arch-datapaths}).

Fig.~\ref{fig:lws-arch} illustrates the LWS architecture, configured with TPU v5e-class compute throughput (393 TOPS INT8), HBM bandwidth (819 GB/s), and memory capacity (16 GiB)~\cite{TPUv5e}. 
LWS consists of 145 compute groups, each containing four MAC clusters, a router, and local SRAM. Each MAC cluster contains 16 MXINT8 MAC tiles, yielding 580 independently schedulable clusters across the chip. 
The compute groups communicate through an on-chip network and share chip-level SRAM and a vector unit. Unlike TPU v5e's four large Matrix Multiplication Units (MXUs), LWS organizes comparable aggregate compute capacity into smaller, independently schedulable clusters that enables finer-grained sharing between Prediction and Learning.

% As shown in Fig.~\ref{fig:lws-arch}(b), LWS employs a \emph{Learning Engine} to manage Learning tasks, weight versions, and updates. 
As shown in Fig.~\ref{fig:lws-arch}(b), LWS employs a distributed \emph{Learning Engine}, implemented as eight slices (one per equipped compute group), to manage Learning tasks, weight versions, and updates.
Each MAC cluster includes lightweight suspension and resumption (R/S) support, allowing Learning kernels to span multiple idle intervals between predictions. 
Eight compute groups additionally include specialized dS/dZ units for softmax gradients and a Stash Quantization (Stash Q) unit for preparing stashed activations and gradients for weight-gradient computation (Section~\ref{sec:arch-datapaths}).

During execution, Prediction operators follow the existing \emph{Control \& Issue} path, while the Learning Engine dispatches ready Learning tasks to available MAC clusters across all compute groups. The Issue Arbiter selects ready Learning tiles based on compute-group availability, while the Prefetch Queue sends DMA read requests for the required operands.
When Prediction requires a MAC cluster occupied by Learning, the Learning task suspends at a safe execution boundary, allowing Prediction to proceed with minimal interference. 
% The Learning Engine applies completed weight gradients to the master weights and prepares updated weight versions, which are then made available to Prediction through the Version Table.
The Weight Update Engine applies completed gradients to the master weights, quantizes the results, and publishes the new version through the Version Table. As Prediction reaches each mutable layer, it acquires the newest published version, waiting only if its staleness exceeds the layer's permitted visibility lag. This allows Prediction to proceed without waiting for the entire backward pass to complete.

\vspace{-10pt}
\subsection{Versioned Weight Execution and Publication}
\label{sec:arch-version}
While Learning produces updated weights, forward and backward computations may still be reading older versions. Updating these weights in place would require waiting for their readers to finish, limiting the overlap between Prediction and Learning. LWS avoids this dependency by separating the publication of new weight versions from the reclamation of old ones. A new weight version is published as soon as it is fully written, while older versions remain available until their readers finish. To coordinate weight publication and bank reuse, LWS includes the Version Table, Publish \& Claim logic, and Weight Update Engine shown in Fig.~\ref{fig:lws-arch}(b).

\textit{Weight-version management.} Each mutable layer maintains one or more weight banks in off-chip DRAM, each storing a complete version of its quantized weights. 
For a permitted visibility lag of $\ell$ chunks, each layer is provisioned with $\ell+1$ weight banks, which are sufficient to avoid bank-allocation stalls in the evaluated workload.
The on-chip Version Table tracks the current published bank, version tags, and a reader count for each bank.
When Prediction reaches a mutable layer, \emph{Control \& Issue} requests the newest published weight version. Each layer's weight-visibility deadline specifies the maximum number of chunks by which its weights may lag behind fully fresh execution. If the newest version satisfies this constraint, the Version Table returns its bank and tag; otherwise, Prediction waits until a sufficiently recent version is published. A bank may have multiple concurrent readers, including forward and backward computations from different chunks. 
Because Learning's backward computation must use the same weight version as Prediction, the Version Table registers separate reader references for both when Prediction acquires the weights. Each reference is released when its corresponding computation finishes. An old bank becomes available for reuse only after all its readers have released their references. 
% Fig.~\ref{fig:lws-bank-timeline}illustrates how these reader references allow multiple weight versions to coexist without delaying publication.

\begin{figure}[!tbp]
    \centering
    \includegraphics[width=.9\columnwidth]{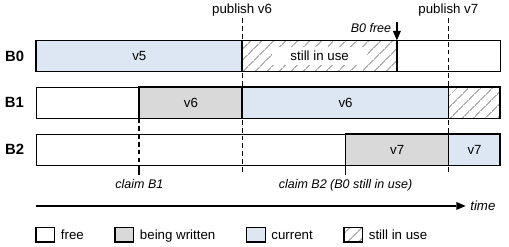}
\caption{\textbf{Weight-bank lifecycle for one mutable layer.}
New weight versions are published without waiting for earlier readers to finish. Older banks remain protected until their readers finish.}
    \label{fig:lws-bank-timeline}
\end{figure}

\textit{Overlapping weight versions.}
Fig.~\ref{fig:lws-bank-timeline} illustrates the weight-bank lifecycle for a mutable layer with a visibility lag of two chunks. Prediction may consume weights up to two updates old, allowing older versions to remain in use while Learning produces newer ones. The example uses three banks, B0--B2. Initially, B0 holds version $v5$. Learning writes $v6$ into B1 while Prediction continues reading $v5$ from B0. Once $v6$ is complete, publishing it makes B1 the current bank without waiting for readers of B0 to finish. When Learning begins producing $v7$, B0 is still in use, so B2 is claimed for the next update. By the time $v7$ is published, B0's readers have finished, making it available for reuse. This allows Learning to publish new versions while every reader keeps the version it acquired.

\begin{figure}[!tbp]
    \centering
    \includegraphics[width=\columnwidth]{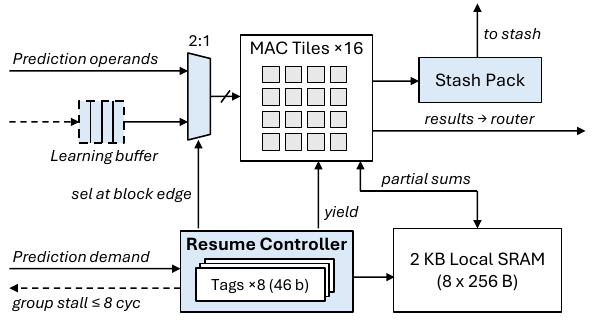}
    \caption{\textbf{Resumable MAC-cluster execution.} Prediction and Learning share the existing MAC tiles through a group-shared operand selector. Learning yields at a reduction-block boundary, saving running partial sums and a compact execution descriptor. The controller later restores this state and resumes Learning without repeating completed blocks.}
    \label{fig:lws-mac}
\end{figure}

\textit{Weight updates and publication.}
The weight banks store quantized weights in the MAC arrays' compute format, used by both Prediction and backward computation. To preserve small weight updates across chunks, LWS also maintains higher-precision master weights that accumulate successive Learning updates without repeated quantization. Only the quantized weights are versioned, while the master weights are updated sequentially.
To produce a new version, the Publish \& Claim logic reserves a destination bank that is neither current nor held by any reader. If no such bank is available, the update waits until one becomes free. For a lag-0 layer, the update overwrites its single bank in place when no computation is reading that layer.

% The DMA streams master weights from DRAM into an on-chip SRAM staging buffer. The Weight Update Engine scales the completed gradients by the learning rate, aligns them for addition to the master weights, and applies the updates.
% The Weight Update Engine supports both stochastic gradient descent (SGD) and AdamW for updating model weights.
% The updated master weights are written back to DRAM, while an on-chip quantizer produces the quantized weights for the destination bank. Once the bank is fully written, the Learning Engine publishes the new version through the Version Table.

The DMA streams master weights from DRAM into an on-chip SRAM staging buffer. The Weight Update Engine processes the completed gradients and updates the master weights using stochastic gradient descent (SGD) or AdamW. The updated master weights are written back to DRAM, while an on-chip quantizer produces the quantized weights for the destination bank. Once the bank is fully written, the Learning Engine publishes the new version through the Version Table.

\begin{figure}[!tbp]
    \centering
    \subfloat[$dS$ unit]{%
        \includegraphics[width=.9\columnwidth]{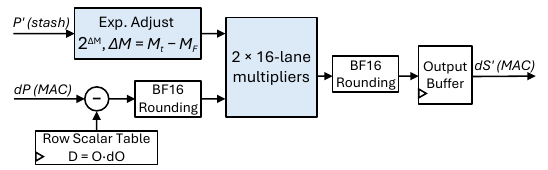}%
        \label{fig:lws-grad-ds}%
    }
    \par
    % \vspace{6pt}
    \subfloat[$dZ$ unit]{%
        \includegraphics[width=.9\columnwidth]{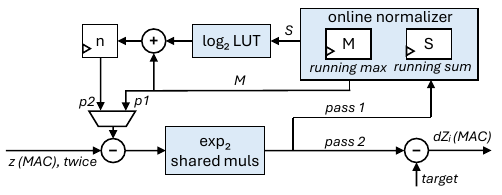}%
        \label{fig:lws-grad-dz}%
    }
    % \caption{\textbf{Dedicated softmax-gradient datapaths.}
    % (a) The attention $dS$ unit adjusts stashed $P'$ by $2^{M_t-M_F}$ to align the tile and final row bounds, then multiplies it by $dP-D$, where $D=r=O^\top dO$. Two 16-lane multiplier banks and BF16 rounding produce the unnormalized gradient $dS'$ in one pass. The remaining row-normalization factor is applied during the subsequent
    % $dQ$ and $dK$ computations.
    % (b) The vocabulary $dZ$ unit streams stashed logits twice through a shared base-2 exponent datapath. Pass~1 maintains the normalization bound $M$ and rescaled exponential sum $S$, then latches $n=M+\log_2 S$. Pass~2 uses these statistics to compute probabilities and subtracts one at the target entry.}
\caption{\textbf{Dedicated softmax-gradient datapaths.}
(a) The $dS$ unit computes attention softmax gradients using stashed probabilities, with normalization deferred to $dQ$ and $dK$.
(b) The $dZ$ unit computes vocabulary softmax gradients in two streaming passes using shared exponent hardware.}
    \label{fig:lws-grad}
\end{figure}

\subsection{Resumable Learning Execution}
\label{sec:arch-resume}

As characterized in Section~\ref{sec:characterization}, Prediction leaves substantial matrix capacity idle, but individual idle windows are shorter than Learning's largest kernels. 
Without suspension and resumption, long Learning kernels cannot fit within individual idle windows, while running them to completion could delay Prediction.
LWS addresses this mismatch through fine-grained suspension and resumption support on existing MAC clusters. This allows Learning kernels to span multiple idle windows while preserving Prediction's priority.

\textit{Prediction-priority arbitration.}
As shown in Fig.~\ref{fig:lws-mac}, each compute group contains a shared operand-selection block that routes Prediction and Learning operands to its four MAC clusters, leaving the underlying MAC datapaths unchanged.
When Prediction requests a cluster executing Learning, the cluster's resume/suspend (R/S) controller initiates a yield at the next reduction-block boundary.
Clusters are lent to Learning only after Prediction completes an output tile and drains its partial sums. While Learning uses the cluster, LWS temporarily buffers the block scales staged for Prediction's next computation and restores them when Prediction reclaims the cluster.

\textit{Block-boundary suspension.}
LWS partitions each Learning tile into reduction blocks of 32 elements, matching the MX block size. Rather than interrupting the MAC pipeline at an arbitrary point, Learning completes the current reduction block before suspending. At this boundary, all products from the block have been accumulated into the running partial sums, allowing the pipeline to drain without preserving its in-flight state. 
% The R/S controller therefore saves only the partial sums, totaling 256\,B per cluster, in the group's local SRAM. 
The R/S controller saves 256\,B of partial sums per suspended tile in local SRAM and supports up to eight suspended tiles per cluster.
A 46-bit descriptor records the layer, weight version, tile range, and reduction-block position needed for resumption.

\textit{Resumption.}
When a cluster becomes available for Learning again, the R/S controller restores the saved partial sums and uses the descriptor to identify the next reduction block and reconstruct its operands. The recorded weight version ensures that Learning resumes using the same weights, which remain protected by the Version Table's reader references. Learning then continues without repeating completed reduction blocks.

% \begin{figure}[!tbp]
%     \centering
%     \includegraphics[width=.9\columnwidth]{figs/dz.pdf}
%     \caption{\textbf{Specialized Learning datapaths.}
%     (a) The attention softmax-gradient unit reuses shared multipliers across two passes.
%        (b) The vocabulary softmax-gradient unit streams the logits twice through one exponent datapath. Pass~1 updates the running maximum $M$ and sum $S$, and $n=M+\log_2 S$ is latched at its end. Pass~2 computes $p_i=2^{z_i-n}$ and subtracts one at the target entry.}
%     \label{fig:lws-grad}
% \end{figure}

\subsection{Learning-Specific Datapaths}
\label{sec:arch-datapaths}

While Learning reuses the existing MAC arrays, backward propagation introduces activation preparation and softmax-gradient operations that would otherwise compete with Prediction for shared resources. LWS defers activation preparation to Learning and adds dedicated softmax-gradient units in a subset of compute groups to reduce this contention.

\textit{Activation stash.}
Weight-gradient computation ($dW=X^\top dY$) accumulates contributions across tokens, requiring stored activations and gradients to be rearranged and quantized into 32-token blocks matching the MXINT8 MAC datapath. 
Performing this preparation during Prediction would require additional gather and quantization hardware in every MAC cluster, increasing forward-path cost.
Instead, a lightweight Stash Pack unit in each cluster (Fig.~\ref{fig:lws-mac}) stores activations in BF16 format in their original layout, with vector-generated activations reaching the stash through shared SRAM. During Learning, Stash Q units in the eight equipped compute groups gather and quantize the stored activations and upstream gradients for MAC processing. Deferring this preparation keeps the more expensive hardware off Prediction's critical path and allows it to be shared across clusters.

\textit{Attention softmax gradient.}
Backpropagation through attention computes gradients with respect to the attention scores. For attention probabilities $P$ and their gradients $dP$,
\begin{equation}
\begin{aligned}
r = \sum_j P_j\,dP_j = O^\top dO,\\
dS_j = P_j(dP_j-r).
\end{aligned}
\label{eq:lws-ds}
\end{equation}
where $O$ is the attention output and $dO$ is its upstream gradient. Following FlashAttention-2~\cite{flashattention2}, computing $r$ from $O^\top dO$ avoids a reduction across the entire attention row.

During Prediction, LWS stores unnormalized attention probabilities $P'$ and their normalization metadata in DRAM. During Learning, the dedicated $dS$ unit (Fig.~\ref{fig:lws-grad-ds}) adjusts the stored probabilities to the final row bound and combines them with MAC-generated $dP$ and the row scalar $r$ to produce unnormalized gradients $dS'$. The remaining normalization is applied during subsequent $dQ$ and $dK$ computations. Unlike FlashAttention-2, which recomputes attention probabilities during backward, LWS reuses the stored values.

\textit{Vocabulary softmax gradient.}
At the output layer, the cross-entropy loss produces a gradient over the vocabulary,
\begin{equation}
dZ_i = p_i-\mathbf{1}[i=t],
\label{eq}
\end{equation}
where $p_i$ is the softmax probability of vocabulary entry $i$ and $t$ is the target token.
The dedicated $dZ$ unit (Fig.~\ref{fig:lws-grad-dz}) uses a two-pass streaming datapath based on online softmax normalization~\cite{Milakov}. The first pass computes normalization statistics, while the second generates probabilities and subtracts the target indicator. Both passes share the exponent datapath, avoiding vocabulary-sized intermediate probability storage and explicit division. Together, the $dS$ and $dZ$ units offload the dominant softmax-gradient operations from the shared vector unit.

% ============================================================
% V. EVALUATION
% ============================================================
% \vspace{-5pt}
\section{Evaluation}
\label{sec:evaluation}

\begin{figure*}[!tbp]
    \centering
    \includegraphics[width=\textwidth]{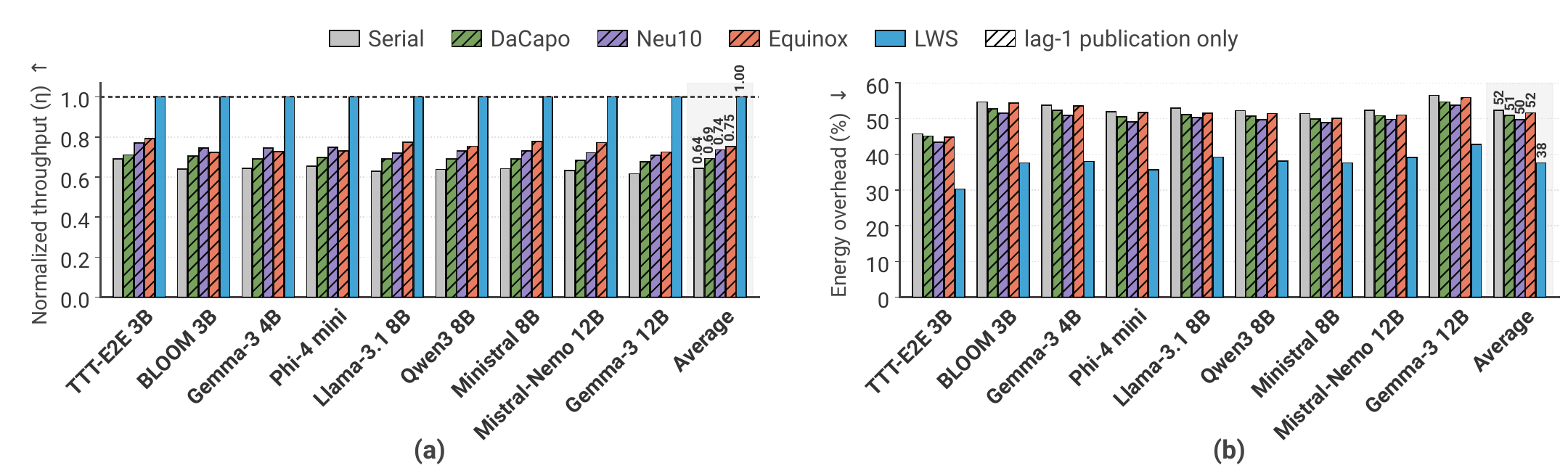}
%     \caption{\textbf{Prediction throughput and Learning energy overhead across nine models.}
% (a) Prediction throughput relative to Prediction-only execution ($\eta=1$ means no slowdown).
% (b) Additional Learning energy per Prediction token, relative to Prediction-only execution.
% Datacenter NPU. Hatched bars are baselines limited to whole-model lag-1 updates. 
% The eight non-TTT-E2E models run the TTT-E2E Learning workload mapped onto their architectures.}
\caption{\textbf{Prediction throughput and Learning energy overhead across nine models on the datacenter NPU.}
(a) Throughput relative to Prediction-only execution ($\eta=1$ indicates no slowdown).
(b) Additional Learning energy per Prediction token relative to Prediction-only execution.
Hatched bars indicate baselines using weights from the preceding chunk while Learning completes.
The eight other models use the TTT-E2E Learning workload mapped onto their architectures.}
    \label{fig:eval-main}
\end{figure*}

\subsection{Setup and Methodology}
\label{sec:eval-method}

% \textit{Workloads.}
% TTT-E2E 3B~\cite{ttte2e} is evaluated on Books3 validation data following TTT-E2E at 32K, 64K, and 128K context lengths. The model has 32 layers with its final eight MLPs mutable, $\mathrm{ITL}(32,8)$, and uses 1{,}024-token chunks with an 8{,}192-token attention window. It serves as the primary workload for both quality and throughput evaluation under the weight-visibility policies characterized in Section~\ref{sec:char-visibility}. 
% To assess throughput across different LLM architectures and model sizes, the TTT-E2E Learning workload is mapped onto eight additional models from 3B to 14B parameters: Llama-3.2 3B, Qwen2.5 3B, Mistral 7B, DeepSeek-LLM 7B, Llama-3.1 8B, Gemma-2 9B, Llama-2 13B, and Qwen2.5 14B. A sweep of 96 TTT-E2E-like configurations (Table~\ref{tab:sweep}) explores the tradeoff between Learning demand and Prediction throughput. LaCT~\cite{TrainingDoneRight} and online LoRA~\cite{online_lora} further extend this analysis across Learning methods.
% All results are for a single long-context sequence during chunked prefill, where ITL learns from the incoming context. Batched serving and autoregressive decoding are outside the scope of this work.
\textit{Workloads.}
TTT-E2E 3B~\cite{ttte2e} is evaluated on the Books3 validation set at context lengths of 32K, 64K, and 128K tokens, following the original TTT-E2E evaluation setup.
The model has 32 layers with its final eight MLPs mutable, $\mathrm{ITL}(32,8)$, and uses 1{,}024-token chunks with an 8{,}192-token attention window. It serves as the primary workload for quality and throughput evaluation under the weight-visibility policies in Section~\ref{sec:char-visibility}.
To assess throughput across LLM architectures and model sizes, the TTT-E2E Learning workload is mapped onto eight additional models: BLOOM 3B~\cite{bloom}, Gemma-3 4B~\cite{gemma3}, Phi-4 mini~\cite{phi4mini}, Llama-3.1 8B~\cite{llama3}, Qwen3 8B~\cite{qwen3}, Ministral 8B~\cite{ministral}, Mistral-Nemo 12B~\cite{mistralnemo}, and Gemma-3 12B~\cite{gemma3}.
Generality across Learning methods is also evaluated using LaCT~\cite{TrainingDoneRight} and online LoRA~\cite{online_lora}.
Performance evaluations focus on a single long-context sequence during chunked prefill, where ITL learns from the incoming context. Batched serving and autoregressive decoding are outside the scope of this work.

% \begin{table}[!tbp]
%     \centering
% % \caption{\textbf{Prior co-execution architectures and ITL requirements.}
% % Nine-model averages for datacenter and edge NPUs (Fig.~\ref{fig:eval-main}).
% % Learning energy overhead is measured per Prediction token relative to Prediction-only execution.}
% \caption{\textbf{Prior co-execution architectures and ITL requirements.}
% Nine-model averages for datacenter and edge NPUs (Fig.~\ref{fig:eval-main}).}
% \label{tab:prior-work}
%     \footnotesize
%     \setlength{\tabcolsep}{2.5pt}
%     \begin{tabular}{@{}lccccc@{}}
%         \toprule
%         & Fine-grained & Vector & Per-layer & & Learning \\
%         Design & sharing & relief & freshness & $\eta$ & energy \\
%         \midrule
%         Serial  & -- & -- & fresh only & 0.61 / 0.63 & 53.2 / 53.7\% \\
%         % V10  & -- & -- & lag 1 & 0.61 / 0.63 & 53.2 / 53.7\% \\
%         DaCapo & static split & -- & lag 1 & 0.69 / 0.70 & 49.8 / 48.4\% \\
%         Neu10   & engine lending & -- & lag 1 & 0.67 / 0.69 & 50.6 / 49.2\% \\
%         Equinox & instruction & -- & lag 1 & 0.73 / 0.75 & 50.3 / 47.5\% \\
%         \textbf{LWS} & \textbf{resumable block} & \textbf{dS/dZ units} & \textbf{per-layer lag} & \textbf{1.00 / 1.00} & \textbf{37.1 / 28.4\%} \\
%         \bottomrule
%     \end{tabular}
% \end{table}

\begin{table}[!tbp]
    \centering
% \caption{\textbf{Prior co-execution architectures and ITL requirements.}
% Nine-model averages for datacenter and edge NPUs (Fig.~\ref{fig:eval-main}).
% Learning energy overhead is measured per Prediction token relative to Prediction-only execution.}
\caption{\textbf{Prior co-execution architectures and ITL requirements.}
Nine-model averages for datacenter and edge NPUs (Fig.~\ref{fig:eval-main}).}
\label{tab:prior-work}
    \footnotesize
    \setlength{\tabcolsep}{2.5pt}
    \begin{tabular}{@{}lccccc@{}}
        \toprule
        & Fine-grained & Vector & Per-layer & & Learning \\
        Design & sharing & relief & freshness & $\eta$ & energy \\
        \midrule
        Serial  & -- & -- & fresh only & 0.64 / 0.63 & 52.4 / 52.1\% \\
        % V10  & -- & -- & lag 1 & 0.61 / 0.63 & 53.2 / 53.7\% \\
        DaCapo & static split & -- & lag 1 & 0.69 / 0.70 & 50.9 / 48.2\% \\
        Neu10   & engine lending & -- & lag 1 & 0.74 / 0.69 & 49.7 / 49.0\% \\
        Equinox & instruction & -- & lag 1 & 0.75 / 0.77 & 51.6 / 47.6\% \\
        \textbf{LWS} & \textbf{resumable block} & \textbf{dS/dZ units} & \textbf{per-layer lag} & \textbf{1.00 / 1.00} & \textbf{37.6 / 29.3\%} \\
        \bottomrule
    \end{tabular}
\end{table}
      
% \textit{Hardware and modeling.}
% LWS is evaluated using an operation-level simulator that models execution dependencies and resource contention between Prediction and Learning across shared MAC, vector, SFU, SRAM, and DRAM resources.
% Matrix-operation latency is modeled using execution cycles measured from RTL simulations of a parallel MXINT8 MAC array.
% The simulator integrates Ramulator~2~\cite{ramulator2} with an HBM2 memory configuration that models DRAM request scheduling, contention and completion. 
% Memory completion events control operand availability and subsequent execution.
% The simulator supports FlashAttention-style tiled attention~\cite{dao2022flashattention} and accounts for all intermediate data traffic.
% SRAM area and energy are modeled using CACTI~\cite{Rajeev2017cacti}, and DRAM energy using DRAMPower~\cite{drampower}.
% Datacenter and edge configurations use compute and memory budgets representative of TPU v5e~\cite{TPUv5e} and Jetson AGX Orin~\cite{JetsonOrin}, respectively. 
% The LWS compute group is implemented in Verilog and verified against MXINT8 reference models.
% Hardware area is evaluated through synthesis in a commercial 16\,nm technology, with power estimated from gate-level analysis. We additionally perform placement of an LWS-equipped compute group to assess physical feasibility.

\textit{Hardware and modeling.}
LWS and all baselines are evaluated using a unified operation-level simulator with consistent workload graphs, hardware budgets, and timing models.
The simulator captures execution dependencies and resource contention between Prediction and Learning across shared MAC, vector, SFU, SRAM, and DRAM resources, including per-layer weight-visibility deadlines.
Matrix-operation latency and LWS preemption overhead are derived from RTL simulations of the parallel MXINT8 MAC datapath.
The simulator integrates Ramulator~2~\cite{ramulator2} for cycle-level DRAM scheduling and contention, using HBM2 for datacenter and LPDDR5 for edge configurations.
Memory completion events control operand availability and subsequent execution.
The simulator supports FlashAttention-style tiled attention~\cite{dao2022flashattention} and accounts for all intermediate data traffic.
All designs are evaluated over multiple chunks to capture steady-state behavior.
SRAM area and energy are evaluated using characterized SRAM macros from the target 16\,nm technology, while DRAM energy is evaluated using DRAMPower~\cite{drampower} with simulated memory activity.
Datacenter and edge configurations use compute and memory budgets representative of TPU v5e~\cite{TPUv5e} and Jetson AGX Orin~\cite{JetsonOrin}, respectively.
The LWS compute group is implemented in Verilog and verified against MXINT8 reference models. 
Hardware area is evaluated using RTL synthesis and characterized SRAM macros in a commercial 16\,nm technology, with logic power evaluated through gate-level analysis.
Placement of an LWS-equipped compute group is also performed to assess physical feasibility.

\textit{Metrics.}
Normalized Prediction throughput $\eta$ is measured relative to Prediction-only execution on the same NPU, where $\eta=1$ indicates no Prediction slowdown. 
% Throughput is measured in steady state, with all required Learning updates completed without accumulating a backlog.
Quality retention measures the fraction of fully fresh Learning's NLL improvement over no Learning, evaluated on 16 input sequences from the Books3 validation set.
The quality-retention results in Section~\ref{sec:characterization} use a BF16 software reference, while the modeled NPU uses MXINT8 matrix computation with higher-precision weight updates. End-to-end validation on TTT-E2E 3B at 128K shows that this implementation introduces less than 0.001 additional NLL compared with the BF16 reference.

\begin{figure}[!tbp]
    \centering
    \includegraphics[width=.9\columnwidth]{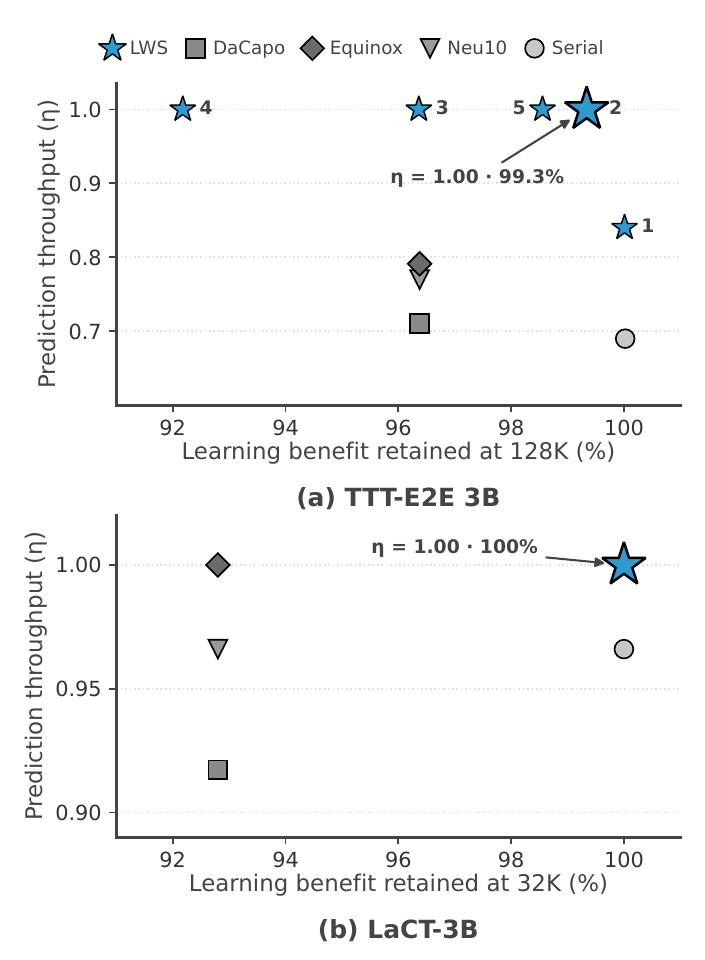}
    \caption{\textbf{Prediction throughput versus retained Learning benefit.}
% (a) TTT-E2E 3B at 128K. LWS policies are (1) fully fresh, (2) \texttt{1111|0000}, (3) uniform lag 1, (4) uniform lag 2, and (5) \texttt{2111|1000}, the selected operating point.
% (b) LaCT-3B at 32K. LWS uses LaCT's native lag-0 freshness, while the baselines use lag 1. }%The panels use different throughput-axis ranges.}
(a) TTT-E2E 3B at 128K. LWS policies are (1) fully fresh, (2) \texttt{1111|0000}, the selected operating point, (3) uniform lag 1, (4) uniform lag 2, and (5) \texttt{2111|1000}.
(b) LaCT-3B at 32K. LWS achieves Prediction-only throughput with fully fresh (lag-0) weights, while the co-execution baselines require whole-model lag 1 to approach or achieve the same throughput.}
    \label{fig:eval-tradeoff}
\end{figure}

\textit{Baselines.}
LWS is compared with serial execution and three prior architectures that support concurrent workloads on a shared accelerator. 
Serial execution follows the TTT-E2E reference schedule, completing each chunk's Prediction, Learning, and weight updates before processing the next chunk.
Neu10~\cite{Neu10} virtualizes cloud NPUs for multiple tenants and lets one tenant borrow the matrix and vector engines that another leaves unused.
DaCapo~\cite{DaCapo} targets continual learning in autonomous systems and statically partitions hardware between inference and retraining. The partition is sized according to Prediction's and Learning's relative MAC operation counts, with each partition receiving a corresponding share of MAC, vector, and SFU resources.
Equinox~\cite{Equinox} trains on an inference accelerator in the cycles that inference leaves idle by interleaving instructions from both.
All three are evaluated on the same underlying NPU organization and resource budget as LWS, while retaining their original sharing mechanisms without LWS-specific hardware.

\textit{Visibility policies.}
Per-layer weight-visibility deadlines trade weight freshness for Prediction throughput. 
% For each model, we evaluate candidate lag policies in the simulator and select the policy used in the main evaluation.
Among policies permitting up to three chunks of lag per layer, LWS uses a low-lag policy that sustains $\eta=1$ with sufficient deadline margin, or the highest-throughput policy found otherwise. 
% For TTT-E2E 3B, this is \texttt{2111|1000}, with an average lag of 0.75 chunks.
For TTT-E2E 3B, this is \texttt{1111|0000}, with an average lag of 0.5 chunks.
% Quality is not evaluated for the eight mapped models, since they were not trained for ITL.
% Unlike LWS, whose hardware-managed per-layer weight versioning (Section~\ref{sec:arch-version}) keeps several versions of each mutable layer and tracks which version every in-flight computation reads, the co-execution baselines do not support per-layer weight-visibility deadlines. They instead keep a single previous copy of the whole model, a whole-model lag of one chunk that lets Prediction proceed while Learning on the preceding chunk completes.
Unlike LWS, the co-execution baselines do not support per-layer weight-visibility deadlines. Instead, they maintain a previous copy of the model's weights, allowing Prediction to use weights from the preceding chunk while Learning completes the updates for the current chunk.

% \textit{End-to-end numerical validation.}
% The numerical precision used in the LWS hardware design is validated against the BF16 reference using TTT-E2E 3B at 128K. With MXINT8 computation and higher-precision weight updates, the selected visibility policy (\texttt{1111|0000}) retains 99.2\% of fully fresh Learning's NLL improvement, close to the 99.3\% measured in the BF16 reference evaluation. The MXINT8 implementation introduces less than 0.001 additional NLL compared with BF16.

%\vspace{-10pt}
\subsection{Comparison with Prior Sharing Mechanisms}

As shown in Fig.~\ref{fig:eval-main}(a), serial execution follows the TTT-E2E reference schedule, completing each chunk's backward pass and weight updates before processing the next chunk. This preserves fully fresh weights but places Learning, including its off-chip gradient and weight traffic, on Prediction's critical path. Across the nine evaluated models, serial execution achieves an average normalized throughput of $\eta=0.64$.

% Prior co-execution architectures partially recover this overhead, reaching average normalized throughput of $\eta=0.69$ for Neu10, $0.74$ for DaCapo, and $0.75$ for Equinox. 
Prior co-execution architectures partially recover this overhead, reaching average normalized throughput of $\eta=0.69$ for DaCapo, $0.74$ for Neu10, and $0.75$ for Equinox.
All three retain a previous copy of the model's weights, allowing Prediction to process the next chunk while Learning completes updates from the preceding chunk.
% Without this overlap, Prediction would have to wait for Learning to finish before entering the mutable layers, substantially reducing throughput.
DaCapo enables concurrent Prediction and Learning by statically partitioning hardware. This reserves dedicated resources for Learning, but reduces the resources available to Prediction even when Learning is idle.
Neu10 dynamically shares resources by lending engines that are not actively used by their assigned workloads. However, during fused attention, matrix engines can remain assigned to Prediction while waiting for dependent vector operations, limiting the idle capacity available for lending.
Equinox interleaves instructions from independent inference and training workloads. However, it cannot suspend and resume partially completed operations, limiting its ability to exploit short, fragmented idle MAC intervals without delaying Prediction.

\label{sec:eval-throughput}
\definecolor{leafEvalSage}{RGB}{226,236,228}
\definecolor{lwsRow}{RGB}{222,236,246}

\begin{figure}[!tbp]
    \centering
    \begin{minipage}[!tbp]{0.49\columnwidth}
        \centering
        \includegraphics[width=\linewidth]{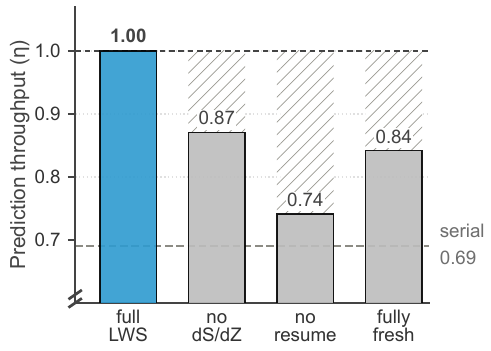}
        {\small (a) Mechanism ablations.\par}
    \end{minipage}
    \hfill
    \begin{minipage}[!tbp]{0.49\columnwidth}
        \centering
        \includegraphics[width=\linewidth]{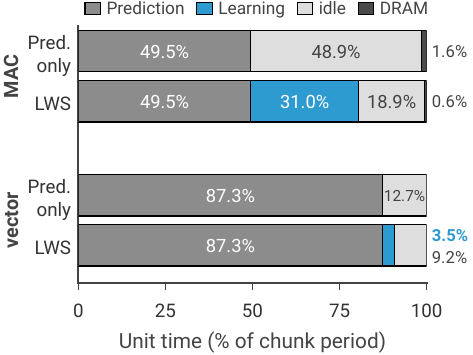}
        {\small (b) Resource utilization.\par}
    \end{minipage}
    \caption{\textbf{Contribution of LWS mechanisms.}
    (a) Normalized throughput under mechanism ablations on TTT-E2E 3B.
    (b) Resource utilization with LWS.}
    \label{fig:eval-ablation}
\end{figure}

\begin{table}[!tbp]
\centering
\caption{Compute-logic area of the datacenter NPU.
Percentages are relative to the baseline compute logic. Shaded rows are added by LWS.}
\label{tab:eval-area}
\begingroup
\footnotesize
\setlength{\tabcolsep}{3pt}
\renewcommand{\arraystretch}{1.05}
\setlength{\aboverulesep}{0pt}
\setlength{\belowrulesep}{0pt}
\setlength{\extrarowheight}{0.8pt}
\begin{tabularx}{\columnwidth}{@{}X r r r r@{}}
\toprule
\textbf{Component} & \textbf{µm$^2$/unit} & \textbf{Count} & \textbf{mm$^2$} & \textbf{\%} \\
\midrule
\textbf{Compute group}\mult{145}                               &          & 145 &       &      \\
\hspace{0.6em}MAC cluster\mult{4}                              & 59{,}069 & 580 & 34.26 & 77.1 \\
\rowcolor{lwsRow}\hspace{1.2em}Resume/suspend\mult{1}          & 374      & 580 & 0.22  & 0.5  \\
\rowcolor{lwsRow}\hspace{1.2em}Stash Pack\mult{1}              & 75       & 580 & 0.04  & 0.1  \\
\hspace{0.6em}Router\mult{1}                                   & 42{,}113 & 145 & 6.11  & 13.8 \\
\rowcolor{lwsRow}\hspace{1.2em}Operand select\mult{1}          & 1{,}655  & 145 & 0.24  & 0.5  \\
\rowcolor{lwsRow}\hspace{1.2em}Learning buffer\mult{1}         & 895      & 145 & 0.13  & 0.3  \\
\rowcolor{lwsRow}\hspace{1.2em}Prediction scale buffer\mult{1} & 1{,}083  & 145 & 0.16  & 0.4  \\
\textbf{Equipped group}\mult{8}                                &          & 8   &       &      \\
\rowcolor{lwsRow}\hspace{0.6em}dS/dZ units\mult{1}             & 37{,}136 & 8   & 0.30  & 0.7  \\
\rowcolor{lwsRow}\hspace{0.6em}Stash Q\mult{1}                 & 9{,}649  & 8   & 0.08  & 0.2  \\
\rowcolor{lwsRow}\hspace{0.6em}Learning Engine slice\mult{1}   & 10{,}688 & 8   & 0.09  & 0.2  \\
\textbf{Vector unit}\mult{1}                                   &          & 1   & 4.04  & 9.1  \\
\rowcolor{lwsRow}\textbf{LWS other}                            & --       & --  & 0.22  & 0.5  \\
\midrule
\textbf{Baseline compute logic}                                &          &     & 44.41 & 100  \\
\rowcolor{leafEvalSage}\textbf{LWS total}                      &          &     & \textbf{1.47} & \textbf{3.3} \\
\bottomrule
\end{tabularx}

\par\smallskip
\begin{minipage}{\columnwidth}
\footnotesize
Post-synthesis at 16\,nm and 1\,GHz. 
%Multipliers give the count per parent block, and each MAC cluster contains 16 MAC tiles. The vector unit has 1{,}024 lanes, matching a TPU-style 8\,$\times$\,128 vector unit. On-chip SRAM (32\,MB, 52.9\,mm$^2$ by CACTI) and I/O are unchanged by LWS and excluded; including SRAM, LWS adds 1.5\%. Percentages are rounded and may not sum exactly. On the edge NPU, LWS adds about 0.37\,mm$^2$ (3.3\%).
\end{minipage}

\endgroup
\end{table}

LWS instead allows Learning to exploit Prediction's idle matrix capacity without reserving a fixed partition or losing partial progress when Prediction needs the hardware. Resumable reduction blocks allow Learning kernels to span short idle intervals and yield when Prediction needs the hardware, preserving their partial results for later execution.
Dedicated softmax-gradient units relieve vector contention, while per-layer weight versioning allows Prediction to proceed without waiting for all Learning updates.
As shown in Fig.~\ref{fig:eval-main}(a), LWS matches Prediction-only throughput ($\eta=1.00$) across all nine models on the datacenter NPU, achieving on average $1.6\times$ the throughput of serial execution and $1.3\times$ that of the strongest prior co-execution baseline, without adding another MAC array.
On TTT-E2E 3B, this corresponds to 7,046 prompt tokens/s on the modeled datacenter NPU, compared with 5,575 for Equinox and 4,861 for serial execution.
Learning's off-chip gradient and weight traffic can also use DRAM bandwidth left idle by Prediction, rather than extending the critical path as in serial execution.
Fig.~\ref{fig:eval-main}(b) shows that LWS also reduces Learning energy overhead. All architectures perform the same essential Learning computation, but slower Prediction extends execution time and increases static energy consumption. By using Prediction's idle compute capacity, LWS limits average Learning energy overhead to 37.6\%, compared with 49.7\% for Neu10 and 52.4\% for serial execution.
The edge NPU demonstrates similar results, with LWS sustaining full Prediction throughput ($\eta=1.00$) compared with $0.77$ for the strongest baseline. LWS also reduces Learning energy overhead to 29.3\%, compared with 47.6--49.0\% for the prior co-execution architectures.

\textit{Throughput--quality tradeoff.}
Fig.~\ref{fig:eval-tradeoff} compares the throughput--quality tradeoffs of LWS weight-visibility policies and prior architectures. %shows how weight-visibility policies allow the same LWS hardware to operate at different throughput--quality points. 
For TTT-E2E, fully fresh Learning preserves the entire quality benefit but limits Prediction throughput to $\eta=0.84$.
Relaxing weight freshness allows more Learning to overlap with Prediction, recovering throughput at the cost of some quality.
However, applying the same delay to every layer is unnecessarily restrictive.
As illustrated by points (2) and (5), per-layer policies achieve more favorable throughput--quality combinations than the corresponding uniform-delay policies (3) and (4).
In particular, \texttt{1111|0000} achieves full Prediction throughput while retaining 99.3\% of Learning's NLL improvement.
In comparison, serial execution preserves fully fresh weights but achieves lower throughput, while DaCapo, Neu10, and Equinox use a one-chunk delay in weight visibility yet still achieve lower throughput than LWS.
% LaCT's lighter Learning workload allows LWS to achieve full Prediction throughput with native lag-0 freshness, retaining 100\% of Learning's quality benefit. Equinox also achieves full throughput, but requires whole-model lag-1 and retains only 92.8\% of the benefit.
LaCT~\cite{TrainingDoneRight} represents a different Learning mechanism, updating weights locally within individual layers rather than backpropagating the next-token loss across multiple layers. LWS also achieves full Prediction throughput for LaCT under its native lag-0 freshness, retaining 100\% of its Learning benefit. Equinox also reaches full Prediction throughput, but with a one-chunk delay in weight visibility, retaining 92.8\% of LaCT's Learning benefit.

\subsection{Ablation and Hardware Cost}
\label{sec:eval-ablation}

Fig.~\ref{fig:eval-ablation}(a) removes one LWS mechanism at a time on TTT-E2E 3B. Without dedicated dS/dZ units, softmax gradients compete with Prediction for the shared vector unit, reducing throughput to $\eta=0.87$. Without resumable reduction blocks, Learning must finish its current tile before yielding to Prediction, reducing throughput to $\eta=0.74$. Finally, requiring fully fresh weights reduces throughput to $\eta=0.84$. Under this constraint, Prediction can still overlap with Learning while traversing the frozen prefix, but may stall upon reaching the first mutable layers, whose weight updates complete last during backpropagation.

\textit{Resource utilization.}
Fig.~\ref{fig:eval-ablation}(b) shows how LWS exploits unused MAC capacity on TTT-E2E 3B. During Prediction-only execution, the MAC array performs no computation for 50.5\% of execution time, with 48.9 percentage points attributed to waiting for softmax operations.
LWS uses these idle periods to execute Learning, increasing MAC utilization by 31.0 percentage points without slowing Prediction.
This realized utilization is higher than the idealized matrix demand in Fig.~\ref{fig:characterization-compute}, which assumes peak MAC throughput. In practice, Learning executes opportunistically on available MAC clusters, where fine-grained sharing and interruptions introduce execution overheads. Nevertheless, Learning still fits within Prediction's unused MAC capacity, leaving 19.5\% of the chunk without active MAC computation.
Unlike the MAC array, the shared vector unit has limited idle capacity. Without dedicated dS/dZ units, Learning would require more than twice the available idle vector time, creating contention with Prediction. By offloading the dominant softmax-gradient operations, the dedicated dS/dZ units reduce Learning's additional vector demand to just 3.5\% of the chunk period. This allows Learning to meet its weight-visibility deadlines without slowing Prediction.
 %, allowing LWS to exploit the available MAC capacity.

\begin{figure}[!tbp]
  \centering
  \includegraphics[width=\linewidth]{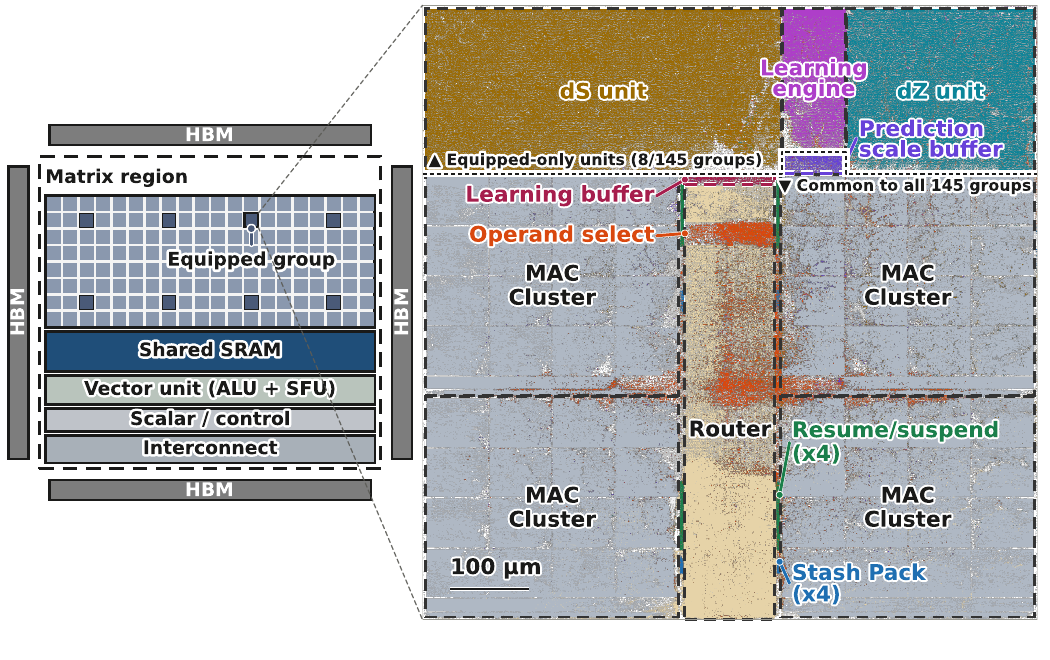}
\caption{\textbf{LWS chip organization and an equipped compute group (16 nm).}
(a) Chip organization. Darker squares mark the eight equipped groups. (b) Placement of an equipped compute group. LWS logic is highlighted, while the four MAC clusters and router retain their original datapaths.}
  \label{fig:floorplan}
\end{figure}

% \textit{Performance sensitivity.}
% We evaluate how changes in hardware resources and available idle matrix capacity affect LWS throughput. Doubling the vector width leaves Prediction's matrix utilization nearly unchanged, with LWS maintaining $\eta=1.00$. Increasing attention matrix activity from 45\% to 55\% gives $\eta=0.998$, while limiting capturable idle matrix time to 91\% gives $\eta=0.985$.
% LWS also meets all per-layer weight-visibility deadlines, with at least 32.4\,ms of steady-state slack on TTT-E2E 3B while keeping outstanding Learning work bounded.

\textit{Hardware implementation and cost.}
Fig.~\ref{fig:floorplan}(a) shows the chip organization, with dS/dZ units, Stash~Q, and Learning Engine slices placed in eight of the 145 compute groups (two of 37 on the edge NPU).
Fig.~\ref{fig:floorplan}(b) shows the placement of an LWS-equipped compute group, where the additional logic surrounds the existing MAC clusters and router without modifying their datapaths.
% The RTL validates group-level Learning execution and preemption. At chip scale, LWS routes Learning tasks and operands over the existing on-chip network, while the simulator models MAC sharing across all compute groups.
As shown in Table~\ref{tab:eval-area}, LWS adds 1.47\,mm$^2$ of synthesized logic to the datacenter NPU and 0.49\,mm$^2$ to the edge NPU, increasing baseline compute-logic area by 3.3\% and 4.3\%, respectively.
Including on-chip SRAM (32\,MB datacenter, 4\,MB edge) at 16\,nm macro density, the added logic accounts for only 1.5\% and 2.7\% of the respective NPU core areas.
The dS/dZ units contribute the largest individual area (0.30\,mm$^2$).
By deferring activation gathering and quantization to Learning and placing Stash~Q only in equipped groups, LWS limits per-cluster additions to a resume/suspend controller and Stash Pack unit.
A standard compute group increases in area by approximately 2\%, compared with 23\% for an equipped group.

% For TTT-E2E 3B, the weight-version banks occupy 624\,MB of DRAM (21\% of the model's quantized weight storage), of which 268\,MB is additional storage introduced by versioning.
% The 1.04\,GB of higher-precision master weights is common to both serial ITL and LWS.
% The simulator accounts for 8.4\,GB per chunk of attention-probability read/write traffic, along with 2.43\,GB of master-weight, residual-state, and version-publication traffic.
% These additional costs fit within the available DRAM bandwidth, allowing LWS to maintain Prediction-only throughput.
% Across the nine evaluated models, each layer holds at most $\ell+1$ live weight versions, with no publication stalls during steady-state execution.

% \textit{Memory overhead.}
% For TTT-E2E 3B under the selected \texttt{1111|0000} policy, the quantized model weights occupy about 3.0\,GB of DRAM. LWS adds 178\,MB for weight versioning, bringing the mutable-layer weight banks to 535\,MB in total. Since only the final eight MLPs are updated, higher-precision master weights are needed only for these layers, requiring another 1.04\,GB shared by both serial ITL and LWS. Together, the weight storage amounts to approximately 4.2\,GB.
% For memory traffic, attention-probability reads and writes account for 8.5\,GB per chunk, while master-weight accesses, residual state, and version publication contribute another 2.43\,GB. This traffic is included in the simulator, and LWS maintains Prediction-only throughput without weight-publication stalls during steady-state execution.

\textit{Memory overhead.}
For TTT-E2E 3B, the quantized model weights occupy about 3.0\,GB, with LWS adding 178\,MB for weight versioning. The higher-precision master weights of the eight mutable MLPs require another 1.04\,GB, required by both serial ITL and LWS that brings the total weight storage to approximately 4.2\,GB.
The simulator also accounts for 8.5\,GB of attention-probability traffic and 2.43\,GB of master-weight, residual-state, and version-publication traffic per chunk. With these costs modeled, LWS maintains Prediction-only throughput without weight-publication stalls during steady-state execution.

\subsection{Generality Across Configurations and Learning Methods}
\label{sec}

% This section evaluates how LWS performs as Learning demand approaches or exceeds Prediction's idle capacity across different model configurations and Learning methods.

\begin{figure}[!tbp]
    \centering
    \includegraphics[width=\columnwidth]{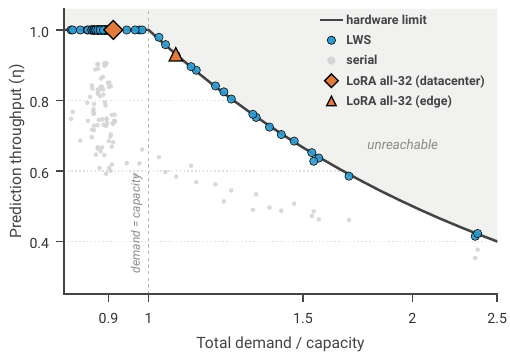}
\caption{\textbf{ITL throughput roofline across 96 configurations.}
Normalized Prediction throughput versus resource demand relative to capacity. The dashed vertical line marks the capacity threshold, while the solid line shows the throughput roofline which LWS reaches in 94 of 96 configurations.}
    \label{fig:eval-generality}
\end{figure}

\textit{ITL configuration sweep.}
Fig.~\ref{fig:eval-generality} evaluates 96 TTT-E2E-like configurations (Table~\ref{tab:sweep}), varying model size, mutable depth, chunk size, optimizer, and NPU configuration. 
% Mutable depth is limited to a quarter of the model's layers, following TTT-E2E~\cite{ttte2e}.
Model size and mutable depth determine backward computation and weight-update volume, while chunk size changes the amortized cost of updates performed once per chunk. AdamW adds optimizer-state traffic and computation relative to TTT-E2E's stochastic gradient descent (SGD), and datacenter and edge NPUs test different hardware resource budgets.
The figure plots normalized Prediction throughput against combined Prediction and Learning resource demand, normalized to hardware capacity over one Prediction-only chunk period. Full Prediction throughput is theoretically achievable when combined demand fits within hardware capacity. When demand exceeds any resource's capacity, that resource imposes a lower throughput ceiling.

\begin{table}[!tbp]
    \centering
    \caption{\textbf{Configurations evaluated in Fig.~\ref{fig:eval-generality}.} The maximum mutable depth is limited to one quarter of each model's layers, following TTT-E2E.}
    \label{tab:sweep}
    \footnotesize
    \begin{tabular}{@{}ll@{}}
        \toprule
        Parameter & Values \\
        \midrule
        Model size & 1B, 3B, 8B \\
        Mutable layers & 2, 4 (1B)\quad 2, 4, 8 (3B, 8B) \\
        Chunk size (tokens) & 512, 1024, 4096 \\
        Inner optimizer & Stochastic Gradient Descent (SGD), AdamW \\
        NPU & Datacenter, edge \\
        \midrule
        Total & 96 configurations \\
        \bottomrule
    \end{tabular}
\end{table}

LWS sustains full Prediction throughput in all 77 configurations whose combined resource demand fits within hardware capacity. Across the 96 configurations, LWS reaches the throughput roofline in 94 cases, including 17 of 19 configurations exceeding capacity. In contrast, serial execution remains substantially below the bound because Learning cannot overlap with Prediction.

% \textit{Online LoRA.}
% We further evaluate LWS on online LoRA~\cite{online_lora}, which updates low-rank adaptation matrices rather than full weight matrices as new chunks arrive. In this configuration, online LoRA has mutable adapters in all 32 transformer layers, compared with TTT-E2E's final eight mutable layers. Despite updating more layers, its combined resource demand remains within hardware capacity. LWS sustains full Prediction throughput, demonstrating that Learning can be hidden even when updates span the entire model. 
\textit{Online LoRA.}
We further evaluate LWS on online LoRA~\cite{online_lora}, which adapts the model to incoming chunks by updating small, low-rank matrices instead of full weight matrices. Unlike TTT-E2E, which updates the final eight layers, this configuration places mutable adapters in all 32 Transformer layers, extending Learning across the entire model. LWS sustains full Prediction throughput ($\eta=1.00$) on the datacenter NPU. On the edge NPU, the combined MAC demand exceeds hardware capacity by 7\%, limiting throughput to $\eta=0.93$, which LWS achieves. 
% These results demonstrate that LWS can reach the hardware throughput limit across different Learning methods, even when available compute capacity is insufficient to fully overlap Learning with Prediction.

% \vspace{-5pt}
\section{Related Work}
\label{sec:related}

\paragraph{Training and inference accelerators}
Prior training accelerators improve efficiency through specialized compute architectures~\cite{tpu,TPUv4,Gaudi,ScaleDeep}, low-precision arithmetic~\cite{rapid,Cambricon-Q}, sparsity~\cite{Procrustes,TensorDash}, and memory optimizations~\cite{GradPIM,PipeLayer,vDNN,Gist,Checkmate}.
Other work reduces on-device training costs through mixed-precision computation~\cite{LNPU}, simplified gradient propagation~\cite{DF-LNPU}, and sparse backward computation~\cite{TinyTraining,PockEngine}.
Ekya~\cite{Ekya} coordinates periodic retraining and inference for video analytics, while PipeDream~\cite{PipeDream} pipelines conventional minibatch training across devices.
Prior inference accelerators have targeted low-latency serving~\cite{Brainwave}, non-GEMM operations~\cite{Tandem}, and fused attention computation~\cite{SystolicAttention}.

% \vspace{-5pt}
\paragraph{Shared accelerators for concurrent workloads}
Prior work improves accelerator utilization through preemption~\cite{PREMA,Tanasic,Chimera,V10}, spatial and temporal partitioning~\cite{Planaria,AIMT,Layerweaver,MoCA,Veltair,DaCapo}, engine lending~\cite{Neu10}, instruction interleaving~\cite{Equinox}, and fast switching~\cite{PipeSwitch,Salus}. These approaches share hardware across concurrent workloads. In contrast, LWS exploits fragmented idle matrix capacity within Prediction to execute its own Learning, using per-layer weight-visibility flexibility to keep Learning off Prediction's critical path.

% \vspace{-5pt}
\section{Conclusion}
\label{sec:conclusion}

Inference-time learning introduces computation that is necessary but not uniformly urgent. Selected layers can tolerate delayed weight visibility, while Learning's matrix work can exploit the idle capacity left by Prediction. This work presents LWS, an inference accelerator architecture that turns this flexibility into hardware. Per-layer weight versioning manages update visibility, resumable execution exploits fragmented idle matrix capacity, and specialized softmax-gradient units relieve vector contention. When Learning fits within Prediction's idle capacity, as on TTT-E2E 3B, LWS matches Prediction-only throughput while preserving 99.3\% of Learning's quality benefit, without an additional compute array. These results suggest that continuous Learning does not require separate training hardware, only a new way to use the compute that inference accelerators leave idle.

% \begin{acks}
% This document is derived from previous conferences, in particular ISCA 2025.
% \end{acks}

% \bibliographystyle{ACM-Reference-Format}
\bibliographystyle{unsrt}
\bibliography{refs}

\end{document}